# A Novel Portable and Wearable Broadband Near-Infrared Spectroscopy Device for In-Vivo Oxygenation and Metabolism Measurements

Musa Talati[1], Frédéric Lange[1], Dimitrios Airantzis[1], Danial Chitnis[1], Temisan Ilukwe[1], Darshana Gopal[1], Paola Pinti[1], Niccole Ranaei-Zamani[2], Olayinka Kowobari[2], Sara Hillman[2], Dimitrios Siassakos[2], Anna David[2], Subhabrata Mitra[2] and Ilias Tachtsidis[1]

[1]Department of Medical Physics and Biomedical Engineering, University College London, London, United Kingdom
[2]Institute for Women's Health, University College London, London, United Kingdom

**Abstract** Broadband NIRS (bNIRS) is an extension of fNIRS that provides the same assessment of oxygenation biomarkers along with a valuable marker for oxygen metabolism at a cellular level, the oxidation state of cytochrome-c-oxidase (oxCCO). bNIRS implements many (100s) NIR wavelengths in the full NIR spectrum to address this and provide insight to tissue energetics. To supply these many wavelengths of light, broadband sources are required, and spectrometers are employed to distinguish power per wavelength. Current multi-channel bNIRS instruments are bulky and only semi-portable due to technological limitations. We propose a design for a bNIRS device that has been miniaturized to allow for portable use. This design leverages the innovations in photonic devices that have created a new line of microspectrometers and broadband NIR high-power LEDs; the Hamamatsu SMD-type spectrometer C14384MA and the Ushio SMBBIR45-1100 LED. This first-of-its-kind device, referred to as microCYRIL (after its two predecessors CYRIL and miniCYRIL), has been developed for oxygenation and metabolism measurements with dual channel operation. To verify functionality, concentration changes in oxygenated ($HbO_2$) and deoxygenated (HHb) haemoglobin and oxCCO were successfully tracked during a cuff-induced venous and arterial occlusion.

## *1. Introduction*

Tissue oxygenation and metabolism measurements are vital for the detection and prognosis of injury and disease. Near-Infrared Spectroscopy (NIRS) provides a non-invasive strategy to measure these through in-vivo quantifications of changes in biomarker concentrations, namely oxygenated and deoxygenated haemoglobin ($HbO_2$ and HHb) and cytochrome-c-oxidase (CCO). NIRS has become a valuable tool for both research and clinical use, with many developments taking functional NIRS (fNIRS) devices from the lab to real-world environments thanks to efforts in portability and wearability. Though, current commercial implementations only provide oxygenation measurements and have limited ability in measuring the metabolic changes important for neurodevelopmental assessments [1].

The oxidisation state of CCO informs on the metabolic function of the cell as the enzyme is the terminal electron acceptor in the electron transport chain, reducing oxygen to form water and allowing adenosine triphosphate (ATP) production. The enzyme contains a copper A redox centre which dominates the absorption spectra of the molecule at NIR wavelengths, with a strong peak in the oxidised form around 830-840nm [2]. The total CCO concentrations themselves do not change over short periods, therefore NIRS is used to provide a marker for changes in concentration of oxidised CCO, [oxCCO]. However, due to its relatively low concentrations when compared to haemoglobin, NIRS faces challenges when measuring the [oxCCO] changes despite its considerably higher specific extinction spectrum. Attenuation variations due to metabolism can be lost within the larger attenuation changes from the $HbO_2$ and HHb, resulting in a chromophore crosstalk; this is when a change in one element produces a spurious measurement of a change in another. These spurious measurements have been avoided by increasing the number of wavelengths in the NIRS range [3], a concept implemented by broadband NIRS (bNIRS).

bNIRS uses many, at times 100s, of NIR wavelengths of light to resolve these slight changes in the absorption spectrum that CCO is responsible for. Compared to haemoglobin oxygenation changes, this can provide a marker for cell activation directly via metabolism [4]. Current bNIRS instruments are bulky and only semi-portable, requiring significant effort to move from room to room. There is an urgent need to bring these up to the state of fNIRS systems, in miniaturisation and portability, to provide a tool that can measure tissue oxygen metabolism in-vivo in a range of environments. Some "mini"-bNIRS devices have been formulated from small-volume spectrometers [5, 6], however, these solutions are still static systems constrained to laboratories and/or connected to the subject via long, and at times restricting, fibres.

Through recent advancements in micro-electronics and photonics, significant changes to bNIRS components have occurred. Particularly, there's been a switch from CCD to CMOS sensors and from Czerny-Turner configurations to varied, integrated spectrometer designs. Coupled with broadband NIR high-power LEDs, these advancements have led to the development of a first-of-its-kind device named microCYRIL (after its predecessors CYRIL [7] and miniCYRIL [6]). This device is designed for oxygenation and metabolism measurements with dual-channel dual-distance operation. To verify functionality, concentration changes in oxygenated (HbO2) and deoxygenated (HHb) haemoglobin and oxCCO were tracked during a cuff-induced venous and arterial occlusion.

## 2. Methods

### 2.1 System Design

| Specification | Item | Description |
|---|---|---|
| Dimensions (mm) | Control Module | 135 x 104 x 48 |
| | Sensor Module | 49 x 135 x 15 |
| | Cable | 1000 |
| Electrical | Supply | USB-C PD: 15V DC |
| | Power requirements | 45W @ 15V DC |
| Battery | Power output (W) | 100 |
| | Capacity (mAh) | 20,000 |
| Single-Board Computer | Device | LattePanda 3 Delta 864 |
| | Processor | Intel Celeron N5105 |
| | Graphics | Intel UHD Graphics |
| | Memory | 8GB LPDDR4 2933MHz |
| | Storage | 64GB eMMC 5.1 |
| | | 32GB Micro-SD |
| | Operating System | Windows 10 IoT 2021 |
| | Video | HDMI 2.0b |
| | USB | 3x USB 3.2 Type-A |
| Microcontroller | Device | Arduino Leonardo |
| | Processor | ATmega32u4-MU |
| | Memory | 2.5kb SRAM, 32kb Flash, 1kb EEPROM |
| | Digital/Analog pins | 20/12 |
| | Communication | UART/I2C/SPI |
| Spectrometer | Device | Hamamatsu C14384MA` |
| | Slit (H x V) (μm) | 15 x 300 |
| | Numerical Aperture | 0.22 |
| | Spectral Range (nm) | 640 to 1050 |
| Light Source | Device | Ushio SMBBIR45A-1100 |
| | Effective Spectral Range (nm) | 400 - 1000 |
|                          | Forward Current (nA)                 | 150                                          |
|                          | Forward Voltage (V)                  | 3.1                                          |
|                          | Rise/Fall time (ns)                  | 90/160                                       |
| Additional Sensors       | Inertial Measurement Unit (MPU6050)  | 6 DOF; 3-axis accelerometer, 3-axis gyroscope |
|                          | Temperature Sensor (LM35)            | Typical accuracy: ±0.25˚C @ 25˚C             |
| Environmental Conditions | Operating Temperature (˚C)           | +5 to +50                                    |
|                          | Storage Temperature (˚C)             | -20 to +70                                   |

**Table 1.** Specification table for the microCYRIL device

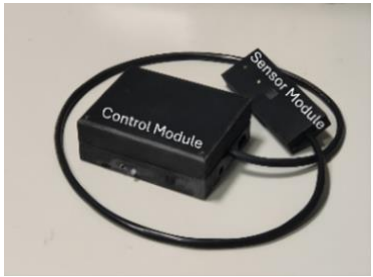

**Fig 1.** Labelled image of the microCYRIL device.

A summary of the device specifications is given in table 1. Attention has been given to the details that are required for the operation of the device. The design is split into two modules: the control module and the sensing module. This setup allows the larger control electronics to be stored in a bag or on a belt away from the monitoring site, while the compact and lightweight sensing module, which includes the source, detector, and environmental sensors, can be held directly against the skin. Figure 1 shows the complete device with both modules labelled.

The device has two channels, created by alternating power to two individual sources at 3 cm and 6 cm source detector separation (SDS), linearly spaced. The light sources are high-power broadband LEDs which provide a high intensity light output in the visible-NIR region (700-1000 nm), which directs light into the skin via a transparent silicone lens. The spectrometer slit is coupled to patient skin via a 90˚ prism, with the detector and source portions of the sensing module separated by a thinner flexible piece of casing material to allow for bending and flexing on the subject area. Additional sensors included in the design are temperature monitoring LM35 sensor and 6-axis movement sensor MPU6050. The sensor module is connected to the control box where the LED power and timing is regulated along with the single board computer that handles spectrometer acquisitions. Here a single push button and RGB LED are available for interface, and a removable SD card for data storage.

## 2.2 Software Design

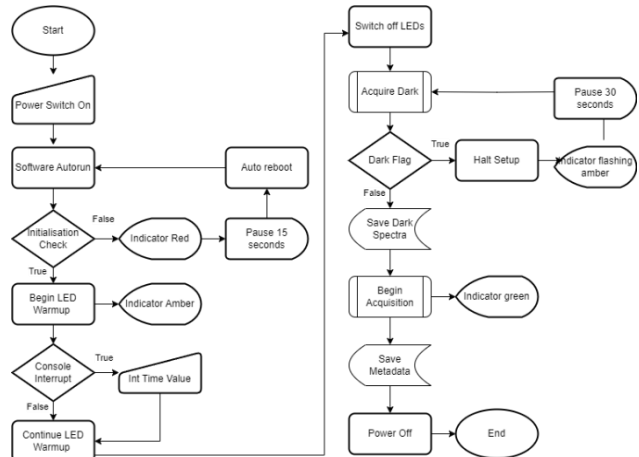

**Fig 2.** microCYRIL high-level software flowchart.

The software for control of the spectrometer acquisitions, the timing sequence of the light sources and the collection of environmental data is programmed as a Windows application that runs automatically when the device is powered on with the power button. As the program steps through each part of the acquisition, an RGB LED on the casing is lit with colors corresponding to current device function. At the end of the acquisition the device can be shut off with a press of the same power button. A flowchart of the high-level software process can be seen in figure 2, the system has been designed so that after an initial configuration of integration times and monitoring duration, it can be run independently of video output with a single button press and stopped similarly.

## 2.3 Cuff Occlusion Protocol

To verify the response of the microCYRIL device when measuring in-vivo changes in oxygenation and metabolism, a venous and arterial cuff occlusion can be performed to induce changes in chromophore concentration and validate the response in the measured spectra.

The spectrometer was set to an integration time of 1000 ms, and the sensing module was placed on the participant's bare forearm along with a blood pressure cuff on the upper arm. Data was collected in the following periods: (A) baseline, (B) venous occlusion, (C) baseline, (D) arterial occlusion, (E) baseline. Venous and arterial occlusion was achieved by applying inflation pressure of 60 mmHg and 230 mmHg respectively. Each period lasted 60 seconds for a total acquisition duration of 5 minutes. To convert the measured spectra into values for change in concentrations, the modified Beer-Lambert law-based UCLn algorithm was applied [8], using 3 cm SDS and a differential path factor (DPF) of 4.16 [9].

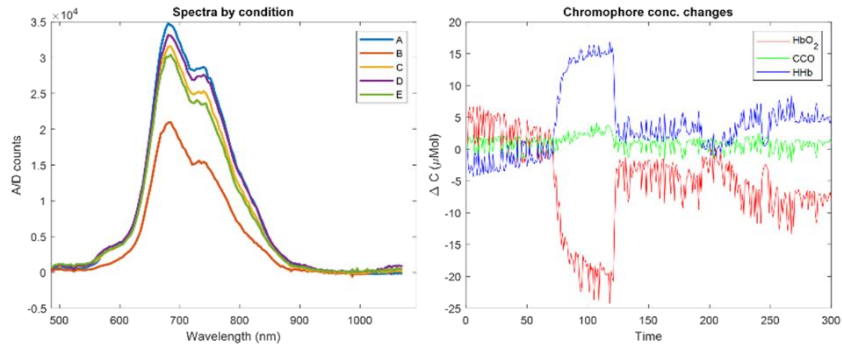

## 3. Results

**Fig 3.** Left: Spectra acquired at a single time point in each period A-E. Right: Changes in concentration of $HbO_2$, HHb and oxCCO (red, blue, and green respectively) in the 5-minute period.

Figure 3 shows expected results of a cuff occlusion validation. With the overall spectral intensity being reduced during each vascular occlusion and recovering in the rest periods, and a large change in the concentrations of the chromophores measured in each occlusion period.

## 4. Conclusion

By leveraging the innovations in miniaturization across photonic components, a novel dual-channel bNIRS system that is portable, fibreless, and wearable has been designed and conceived and demonstrates promising results in the cuff occlusion validation. This device can be used for oxygenation and metabolism tissue measurements with capacity for non-restrictive testing, i.e. in real world environments without limiting movement.